\documentclass[pra,twocolumn,superscriptaddress,showpacs,floatfix]{revtex4}

\usepackage{graphics}           
\usepackage{bm}                 
\usepackage{amsmath}            
\usepackage{amsfonts}
\usepackage{amssymb}
\usepackage{latexsym}           

\begin{document}

\title{Complementarity and quantum walks}

\author{Viv Kendon}
\email{V.Kendon@leeds.ac.uk}
\affiliation{QOLS, Optics Section, Blackett Laboratory, Imperial College,
	London, SW7 2BW, United Kingdom.}
\affiliation{School of Physics and Astronomy, University of Leeds,
	Leeds, LS2 9JT, United Kingdom.}
\author{Barry C.\ Sanders}
\affiliation{Institute for Quantum Information Science,
University of Calgary, Alberta T2N 1N4, Canada}
\affiliation{Centre for Quantum Computer Technology, Macquarie University,
	Sydney, New South Wales 2109, Australia}

\date{April 7, 2004, revised August 20, 2004 and November 9th, 2004}

\begin{abstract}

We show that quantum walks interpolate between a coherent `wave walk'
and a random walk depending on how strongly the walker's coin state
is measured; i.e., the quantum 
walk exhibits the quintessentially quantum property of complementarity,
which is manifested as a trade-off between knowledge of which path the
walker takes vs the sharpness of the interference pattern.
A physical implementation of a quantum walk (the quantum quincunx)
should thus have an identifiable walker and the capacity to demonstrate
the interpolation between wave walk and random walk depending on the
strength of measurement.
\end{abstract}



\maketitle

\section{Introduction}

Random walks are essential to physics as stochastic phenomena,
to mathematics as Wiener processes, and to 
computer science for algorithms.
The quantum walk, both in its continuous~\cite{farhi98a} and in its
discrete~\cite{watrous98a,ambainis01a,aharonov00a}
incarnations, is receiving significant attention because it is
a natural generalization of random walks to quantum systems,
because quantum walks may be physically implemented by
quincunxes~\cite{travaglione01a,sanders02a,dur02a},
and because quantum walks could provide a basis for future quantum
algorithms~\cite{childs02a,shenvi02a,ambainis04a}.  However, an
identifiable benefit of the quantum walk, namely the 
enhancement of spreading over its classical counterpart, is a 
wave phenomenon
which has been realized interferometrically
in an optical quincunx~\cite{bouwmeester99a},
and proposed in other settings \cite{knight03a,knight03b,jeong03a}.
The realization of aspects of the
quantum walk in a classical optics setting has raised the question of what
exactly is `quantum' about the quantum walk.
We resolve this issue of comparing and contrasting the deterministic wave 
walk vs a genuine quantum walk by properly accounting for the role of
complementarity~\cite{bohr28a,bohr28b,bohr50a}.

Although complementarity has been at the heart of quantum mechanics since the
dawn of the subject~\cite{bohr28a,bohr28b}, studies of complementarity often focus on simple,
illustrative cases such as two-slit interference~\cite{wootters79a} and
two-channel interferometry~\cite{Sanders89a,englert96}; we
significantly expand the field by providing an
analysis of complementarity for general graphs.  To incorporate complementarity
into quantum walks, we extend from the typical view of a quantum
walk defined as unitary local transition rules over the Hilbert space
for the vertices of the graph and the states of the walker's coin by allowing
a measurement process, either on the transitions, or on the coin outcomes,
or both. The measurement is performed by entangling the walker or coin to 
ancillary degrees of freedom, with the strength of coupling to the ancilla
determining whether the quantum walk is coherent
(no coupling to ancilla yielding the unitary quantum walk)
or random (strong coupling that yields full information 
on the walker's path). 

We then define a quantum quincunx as a physical implementation of a
quantum walk,
which must have an identifiable walker and interpolate between a random
walk and a unitary quantum walk as the measurement strength is varied.
More precisely, we require a quantum quincunx to have
(i)~a single walker,
(ii)~a measurement process that can be employed to acquire varying degrees
of knowledge about the path of the walker, and
(iii)~an identifiable interference phenomenon whose deterioration can be
linked to the acquisition of knowledge about the walker's path.
This view on complementarity (one walker and a trade-off between which-path vs
interference) follows the information-theoretic perspective of complementarity
introduced by Wootters and Zurek~\cite{wootters79a}.

We proceed as follows: first we provide background on 
complementarity in Sec.~\ref{sec:complementarity}
and then introduce our notation for
general graphs, recalling the definition of a classical random walk on
such graphs, in Sec.~\ref{sec:gengraph}.
This is followed in Sec.~\ref{sec:qwunitary} by the definition
in our notation of a unitary quantum walk on a general graph.
In Sec.~\ref{sec:qwnonu} we extend the definition to include
partial measurements of the quantum walker.
An example of the walk on a $N$-cycle is given in Sec.~\ref{sec:qwcycle}.
In Sec.~\ref{sec:measure} we describe in detail how to 
perform a measurement of the path taken by the walker by measuring the state
of the walker's coin, followed by a general treatment
of quantum walks with nonunitary evolutions in Sec.~\ref{sec:nonunitary}.
In Sec.~\ref{sec:optqquincunx} we discuss how the wave
walk relates to quantum walks, and in Sec.~\ref{sec:conc} we summarize our
results. 

\section{Complementarity for quantum walks}
\label{sec:complementarity}

In its original formulation~\cite{bohr28a,bohr28b},
complementarity is the principle that one classical description
of a system, which explains certain phenomena for a quantum system,
is incompatible with the simultaneous use of another classical
description used to explain other phenomena.
In simpler terms, a quantum system can exhibit different, incompatible
properties that are each manifested under different circumstances.
The most well-known example of complementarity is wave-particle duality:
objects such as single electrons or single photons can be described as
being corpuscular (particle-like) under some circumstances
(when the phenomenon being studied, such as particle
detection, can be explained by describing these objects
as localized, indivisible particles) and undular (wave-like)
under other circumstances (when the phenomenon being studied,
such as interferometry, can be explained by regarding
the objects as extended, interfering waves).
The attributes of waves and particles make these two descriptions
mutually incompatible, yet electrons or photons can be made to
exhibit the features of both these two incompatible descriptions
depending on how they are observed.

Complementarity is at the heart of quantum mechanics.
Electrons and photons are described by quantum theory in order
that these mutually incompatible descriptions can be reconciled.
Quantum mechanics provides a unified framework for describing quantum
systems that can be corpuscular or wave like under different circumstances.
Systems are in fact regarded as quantum when complementarity is manifest.

Although the early descriptions of complementarity concerned mutually
incompatible measurements, Wootters and Zurek presented an
information-theoretic description of complementarity,
which elucidates that complementarity can be quantified
as a trade-off between knowledge of which way each particle
goes vs the sharpness of the interference pattern obtained via
repeated preparations and measurements~\cite{wootters79a}.
This trade-off between corpuscular and undular behavior has been examined
in detail for photons both in a theoretical context using a photon
number quantum nondemolition measurement~\cite{Sanders89a} and
experimentally using non-deterministic linear optical gates~\cite{pryde04a}.

While complementarity has been well studied in quantum physics,
its role in identifying the `quantumness' of quantum information
tasks has not been explored.
Recent controversy over what is `quantum' about quantum walks
motivates us to examine the role of complementarity in this context.
The controversy over the quantum walk is exemplified by the statement by Knight
et al in the abstract of their paper on ``Quantum walk on the line as an
interference phenomenon'' that, ``the coined quantum walk on a line can be 
understood as an interference phenomenon, can be classically implemented,
and indeed already has been''~\cite{knight03a}.
In their conclusions, they state that they have 
``shown that the [quantum walk] along a line can be simulated in a
purely classical implementation, involving nothing more than wave
interference of electromagnetic fields.''
Their work shows that the quantumness of the coin,
which is a spin-$\tfrac{1}{2}$ particle for the quantum walk on the line,
and its possible entanglement with the walker's path,
do not by themselves make the quantum walk `quantum'.
The question then arises whether this reasoning
is sufficient to claim as they do, that the quantum walk is
purely a wave phenomenon that
``can be  simulated \ldots [by] wave interference of electromagnetic fields''.

Our position is that the quantum walk may indeed be implemented by an
optical system, but not by one that is strictly described by classical
electromagnetic theory.
The optical quincunx of Bouwmeester et al~\cite{bouwmeester99a} certainly
displays the interference features of the quantum walk on the line,
but the quantumness of the quantum walk must connect two seemingly
incompatible descriptions: there is a single walker at a time who can opt
for different paths that interfere with each other,
and the acquisition of information about the path destroys the
interference and restores the classical walk.
An experiment that observes one phenomenon of the quantum walk,
the interference, is really only observing a `wave walk';
we will show that a \emph{quantum} optical quincunx
can identify that there has been one walker,
learn about the path, and show that interference
diminishes as path information is obtained.

\section{General graphs}
\label{sec:gengraph}

\begin{figure}
    \resizebox{0.7\columnwidth}{!}{\includegraphics{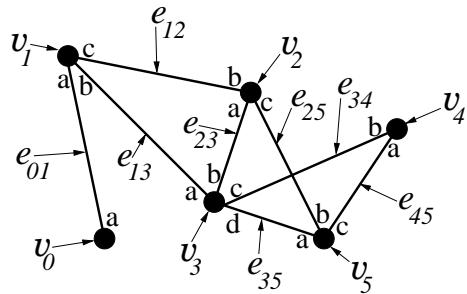}}
    \caption{An example of a general graph, with $N=6$ vertices
            labeled $v_0,v_1\dots v_5$, and eight edges labeled
            $e_{01}\dots e_{45}$. A labeling of the possible choices
            of paths from each vertex is given by the
            letters \{a,b,c,d\}, the
            degree of this graph being four.}
    \label{fig:gengraphcoin}
\end{figure}

We cast our discussion of quantum walks in a very general setting where
the walk takes place on a general graph $G(V,E)$ with
\begin{equation}
V=\{v_j ; j\in\mathbb{Z}_N\}
\end{equation}
the set of vertices, and $E=\{e_{jj'}\}$ the set of edges,
where $e_{jj'}$ connects vertices $v_j$ and $v_{j'}$,
as shown in Fig.~\ref{fig:gengraphcoin}.
The number of edges adjoining vertex~$v_j$ is $d_j$
and the degree of the graph is $d=\text{max}_j d_j$.
The complexity of the graph is associated with $|V|=N$, and
the degree~$d$ is constant as $N$ varies.
For simplicity, we assume \textit{undirected} edges ($e_{jj'} \equiv e_{j'j}$),
and at most one edge between any pair of vertices ($e_{jj'}$ is unique).

In a random walk, the walker's choice of which edge to follow from
a given vertex is random  (for which we include a $d$-dimensional coin),
but this choice of edges can be weighted (some edges are preferred over others)
or biased (the coin outcomes are not uniformly distributed over all choices).
In general the nature of the coin is correlated with the vertex
from which the coin is tossed, e.g.~the coin
is $d_j$-sided at vertex~$v_j$ or the bias of the coin may be $j$-dependent.

\section{Unitary quantum walk}
\label{sec:qwunitary}

In contrast to the classical random walk,
the quantum walk permits the walker to follow all edges
in a superposition state, essentially as Feynman paths through the graph.
We first describe the unitary quantum walk, and then generalize to
include measurement of the walker's progress.
The unitary quantum walk is deterministic,
the walker's wavefunction is in the Hilbert space
$\mathcal{H}_\text{vc}$
which contains the $N$-dimensional Hilbert space
\begin{equation}
	\mathcal{H}_\text{v}=\text{span}\{| j\rangle_\text{v}:j\in\mathbb{Z}_N,
		\!_\text{v}\langle j|j' \rangle_\text{v}=\delta_{jj'}\}
		\subset \mathcal{H}_\text{vc}
\label{Hv:span}
\end{equation}
of vertex states.
For a coin we have a $d$-dimensional Hilbert space
\begin{equation}
	\mathcal{H}_\text{c}=\text{span}\{|k\rangle_\text{c}:k\in\mathbb{Z}_d\}
	\;\;\text{and}\;\;	\!_\text{c}\langle k|k' 
\rangle_\text{c}=\delta_{kk'}\}
\label{Hc:span}
\end{equation}
where~$d$ is the degree of the graph.
The basis states of $\mathcal{H}_\text{vc}$ are given by
\begin{equation}
	\mathcal{B}_\text{vc}=\{|j,k\rangle \equiv |j\rangle_\text{v}|k\rangle_\text{c} ;
		 j\in\mathbb{Z}_N,k\in\mathbb{Z}_{d}\}
\label{basis:vc}
\end{equation}
with cardinality $Nd$.
For a basis state~$|j,k\rangle$,
the index~$j$ identifies the vertex number and $k$ the $k^\text{th}$
state of the coin.
For an edge $e_{jj'}$ we associate the coin state $k$ with the 
edge at $v_j$, and the coin state ${k'}$ with the other end of
the edge at $v_{j'}$.  The values of $k$ and ${k'}$ are
arbitrary but fixed throughout the quantum walk,
to ensure the quantum walker traverses the the graph in a consistent manner.
We define the mapping
\begin{equation}
        \zeta:\mathbb{Z}_N\times\mathbb{Z}_{d}
                \rightarrow\mathbb{Z}_N\times\mathbb{Z}_{d}:
		(j,k)\mapsto\zeta(j,k)=(j',k'),
\label{mapping}
\end{equation}
where $(j,k)$ and $(j',k')$ label each end of $e_{jj'}$.

The quantum walk undergoes a unitary evolution by repetition of two
steps: a coin toss and a conditional swap.
The coin operator 
\begin{equation}
	C:\mathcal{H}_\text{vc}\rightarrow\mathcal{H}_\text{vc}:|j,k\rangle\mapsto
		\sum_{\tilde{k}\in\mathbb{Z}_{d}}C_{k\tilde{k}}^j|j,\tilde{k}\rangle_\text{c}
\label{C:transformation}
\end{equation}
is a block diagonal matrix with each block labeled by $j$.
The $j$-dependence of the coin matrix allows sufficient freedom in
the quantum walk dynamics for
the quantum coherence properties of the coin to vary between vertices,
for vertices to act as origins and endpoints,
and for vertices to have different degrees from each other.
If $v_j$ has degree $d_j<d$, we require $C_{k\tilde{k}}^j=0$ for all
$\tilde{k}$ values not used to label an edge at $v_j$.
This restricts the coin operator so it only produces states that
have a valid mapping under $\zeta$.

If 
\begin{equation}
	C_{k\tilde{k}}^j=C_{k\tilde{k}}^{j'}\forall j,j',
\end{equation}
we have the special case of a fixed degree graph where
the coin operator is identical for all vertices.
As examples, a two-sided coin has been employed in 
analyses of the quantum walk on the line \cite{ambainis01a}
and quantum walk on the cycle~\cite{aharonov00a}, and a multi-sided coin
for quantum walks on regular lattices in higher
dimensions~\cite{moore01a,mackay01a}.

The unitary conditional swap operator is given by
\begin{equation}
	S:\mathcal{H}_\text{vc}\rightarrow\mathcal{H}_\text{vc}:
		|j,k\rangle\mapsto|j',k'\rangle~,
\label{eq:qswap}
\end{equation}
which updates the position of the walker and the coin state
according to the mapping $\zeta$ in Eq.~(\ref{mapping}), i.~e., moves 
the walker and coin to the vertex $v_{j'}$ along edge $e_{jj'}$.
We note that, by our stipulation that $k$ and ${k'}$ label
opposite ends of $e_{jj'}$, it follows that $S=S^{-1}$, and is thus unitary
as required for quantum evolution.
The sequence of a coin flip and a conditional swap is a transition
over the unit time step, which we denote by unitary $T=SC$.

This formulation of a (pure state) coined quantum walk on a general
graph first appeared in a different form due to Watrous \cite{watrous98a},
and is also described by Ambainis \cite{ambainis04a}.  The interferometric
scheme of Hillery et al \cite{hillery03a} is also equivalent.
The astute reader will have noted that there is thus far nothing random 
in the dynamics of a unitary quantum walk, it being a perfectly 
deterministic, reversible unitary evolution.  Randomness does arise
if one measures the position of the walker after a number of time steps,
when the walker will be found on one of the vertices with a
probability given by the squared modulus of the walker's wavefunction
over the graph.

\section{Nonunitary quantum walk}
\label{sec:qwnonu}

We now generalize to include measurements as part of each step of the
quantum walk dynamics (rather than only after the final step).
Quantum walks with measurements were first considered
by Aharonov et al~\cite{aharonov92a} though with different
motivations from ours.
The inclusion of measurements requires nonunitary evolution,
therefore we introduce density matrices to describe the walker's state.
The (time-dependent) density operator
\begin{equation}
	\rho = \sum_{j,k}\sum_{j',k'}\rho_{jk}^{j'k'}|j,k\rangle\langle j',k'|,
\end{equation}
is a positive ($\rho=\rho^\dagger$ with positive real spectrum),
unit-trace, bounded linear operator on $\mathcal{H}_\text{vc}$,
in the basis $\mathcal{B}_\text{vc}$, Eq.~(\ref{basis:vc}).
The state is pure iff $\rho^2=\rho$.
A typical initial condition is $\rho(t=0)=|0,0\rangle\langle 0,0|$
corresponding to the walker starting at vertex $v_0$ carrying a coin in the $0$ state.

In general the density operator is mapped to a new density operator via
a completely positive, or CP, map
\begin{equation}
\mathcal{T}:\rho\mapsto\mathcal{T}\rho
\end{equation}
The CP map $\mathcal T$ performs 
one coin flip and the conditional swap over one time step.
As a CP map,
\begin{equation}
	\mathcal{T}\rho=\sum_{i\in\Theta} T_i^\dagger \rho T_i,
	\text{\null\hspace{1em}\null}
	\sum_{i\in\Theta} T_i^\dagger T_i = \openone,
\label{T:decomposition}
\end{equation}
with $i$ an index of nonunitary evolutionary `instances'
and $T_i$ the corresponding
Kraus operator.
These instances may be discerned by a measurement record, with $i$ the
record index.
The cardinality of $\Theta$ can be finite, countably infinite, or even uncountable.
In the case of unitary evolution, $\Theta$ has a cardinality of one, so there is a unique,
unitary $T$ for which
$\rho\mapsto\mathcal{T}\rho = T \rho T^\dagger$.
Unitary quantum walk evolution can thus be expressed as
\begin{equation}
	\rho(t) = \mathcal{T}^t \rho(0),\; \mathcal{T}\equiv\mathcal{SC},
		\; \mathcal{SC}\rho\equiv SC\rho C^\dagger S^\dagger~,
\label{T:unitary}
\end{equation}
where, for the discrete time walk, we assume $t\in\mathbb{Z}$.
Thus, for the unitary walk, the transition is given by $T=SC$,
and the nonunitary walk can be understood as a collection,
or sum, of instances of nonunitary coin flips, with randomness,
followed by a conditional swap.

\section{Walk on the cycle}
\label{sec:qwcycle}

\begin{figure}
    \resizebox{0.7\columnwidth}{!}{\includegraphics{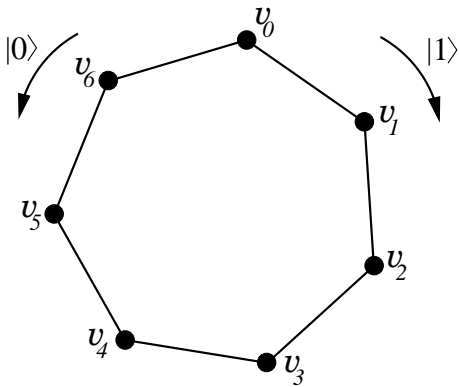}}
    \caption{An example of a cycle graph, with $N=7$ vertices
            labeled $v_0,v_1\dots v_6$, and seven edges.
	    Arrows indicate the direction of travel for coin state
	    $|0\rangle$ (anticlockwise) and $|1\rangle$ (clockwise).
	    The cyclic labeling of the vertices does away with the need
	    to label the edges.}
    \label{fig:cycle}
\end{figure}

At this stage application to a well known example is helpful,
and we consider the quantum walk on a cycle, see Fig.~\ref{fig:cycle}.
The quantum walk on the cycle has the advantages that 
$\mathcal{H}_\text{v}$ is finite-dimensional
(as opposed to the walk on the line, for example,
which has infinitely many vertices, hence an infinite-dimensional 
Hilbert space), $d=2$ for all vertices,
which is the smallest nontrivial degree, and the quantum walk
on the cycle may be experimentally achievable~\cite{travaglione01a,sanders02a}.
The Hilbert space for a two-sided coin is given by 
$\mathcal{H}_\text{c}=\mathcal{H}_2$ for
\begin{equation}
	\mathcal{H}_2 = \text{span}\{|0\rangle,|1\rangle\}~.
\label{H2}
\end{equation}
The Hilbert space for the graph+coin is given by
\begin{equation}
\mathcal{H}_\text{vc}=\mathcal{H}_\text{v}\otimes\mathcal{H}_\text{c}.
\end{equation}

We can choose all blocks of the coin matrix to be identical
$2 \times 2$ matrices, in this case the unbiased two-sided coin operator 
is given by a generalized Hadamard transformation~\cite{mackay01a}
\begin{equation}
	H_\varphi\equiv\tfrac{1}{\sqrt{2}}
		\begin{pmatrix}1&-\text{i}\text{e}^{\text{i}\varphi}\\ 
		\text{i}\text{e}^{\text{-i}\varphi}&-1\end{pmatrix},
\label{H:phi}
\end{equation}
with a free phase degree of freedom~$\varphi$
(typically $\varphi=\pi/2$).
Assuming that the vertices are labeled in sequence around the cycle,
we can employ the simplicity of assigning $|0\rangle$ to moving from 
$v_j$ to $v_{j+1}$ and $|1\rangle$ to moving from $v_j$ to $v_{j-1}$
(rather than labeling each end of the edges).
The conditional swap operation becomes
\begin{equation}
	S =\sum_{\varepsilon\in\{0,1\}}\sum_{j\in\mathbb{Z}_N} |j-(-1)^\varepsilon\bmod N\rangle
			\langle j|\otimes|\varepsilon\rangle\langle\varepsilon|
\end{equation}
yielding the transformation
$S|j,\varepsilon\rangle=|j-(-1)^\varepsilon,\varepsilon\rangle$.

The quantum walk on the cycle has been well-studied:
it mixes faster than a classical random walk, 
both for unitary quantum walks \cite{aharonov00a},
and when a small amount of decoherence is applied in the form of
non-unitary random measurements of the coin and/or the walker's
position \cite{kendon02c}.

\section{Measurements of the coin}
\label{sec:measure}

Now that we have the machinery in place to describe nonunitary quantum walks
on general graphs, we can return to the question of complementarity and
consider how to track the path of the walker.
We will explain this in detail for a two-sided coin such as the one
used for the walk on a cycle described in the previous section.
Suppose that measurements of the coin state are performed
after each coin flip, in the coin state basis $\{|0\rangle,|1\rangle\}$.
This measurement can be performed by adding an ancilla that serves as the
coin meter, and the meter state becomes correlated by interacting with the coin.
The Hilbert space for the ancilla is of the same dimension as the Hilbert space
for the coin, so the meter's Hilbert space is
$\mathcal{H}_\text{m}=\mathcal{H}_2$, given by Eq.~(\ref{H2}).
Letting
\begin{align}
	\sigma_+ \equiv |1\rangle_\text{c} \langle 0|=\sigma_-^\dagger,
	&\;\tau_+ \equiv |1\rangle_\text{m} \langle 0|=\tau_-^\dagger,	\\
	2\sigma_x = \sigma_+ + \sigma_-, &\;2\text{i}\sigma_y=\sigma_--\sigma_+~,	\\
	2\sigma_z=[\sigma_+,\sigma_-], &\;2\tau_z=[\tau_+,\tau_-],
\label{sigmatau}
\end{align}
the meter couples to the coin via the interaction~\cite{berry03a}
\begin{align}
W(\beta) =&
	\left(\text{e}^{\text{i}\frac{\pi}{4}\sigma_y}\otimes\text{e}^{-\text{i}\frac{\pi}{4}\tau_x}\right)
	\text{e}^{-\text{i}\beta\frac{\pi}{4}\sigma_x\otimes\tau_x}
	\left(\text{e}^{-\text{i}\frac{\pi}{4}\sigma_y}\otimes
	\text{e}^{-\text{i}\frac{\pi}{4}\tau_z}\right)
		\nonumber	\\
	&\times	\left(\openone\otimes\text{e}^{-\text{i}\pi(1-\beta)\tau_y/4}\right)
		\left(\openone\otimes\text{e}^{-\text{i}\pi\tau_z/4}\right).
\label{eq:ndop}
\end{align}
The uncoupled case corresponds to $W(0)$, and $W(1)$ corresponds to the strong
coupling limit (a controlled-NOT operation) with resultant sharp measurements.
The interpolation between the limits is achieved by allowing
$\beta$ to vary continuously over the domain $[0,1]$.

To perform a measurement, we first prepare the meter in the
$|0\rangle_\text{m}$ state, then allow it to interact with the coin, which is in
an arbitrary qubit state $\gamma|0\rangle_\text{c}+\eta|1\rangle_\text{c}$,
\begin{align}
	W(\beta) (&\gamma|0\rangle_\text{c}+\eta|1\rangle_\text{c})|0\rangle_\text{m}
		= \gamma |00\rangle_\text{cm} 
			\nonumber	\\
		&+ \eta|1\rangle_\text{c}
		(\cos\beta\pi/2|0\rangle_\text{m}+\sin\beta\pi/2|1\rangle_\text{m}).
\end{align}
Tracing over the meter state yields the $2 \times 2$ coin density matrix transformation
\begin{align}
	\text{Tr}_\text{m}&\{W(\beta)\left[\begin{pmatrix}|\gamma|^2 & \gamma\eta^* \\
		\gamma^* \eta & |\eta|^2 \end{pmatrix} \otimes \begin{pmatrix}
		1&0\\0&0\end{pmatrix}\right]W^\dagger(\beta)\}
			\nonumber	\\
		&=\begin{pmatrix}|\gamma|^2 & \gamma\eta^* \cos\beta\pi/2\\
		\gamma^* \eta\cos\beta\pi/2 & |\eta|^2 \end{pmatrix}~.
\label{eq:W}
\end{align}
The coin values correspond to the pointer basis~\cite{zurek81a,zurek82a},
with the degree of decoherence depending on $\beta$.

These measurements of the coin provide information on path
choices that the walker makes.  
If the walker's initial state is localized at one known vertex, then perfect
coin state measurements suffice to inform the observer as to the exact trajectory of
the walker. The parameter~$\beta$ allows the observer to acquire as much or as
little information as desired,
thereby allowing the degree of knowledge of which 
trajectory the walker followed and the complementary
degree of coherence reduction to be controlled.
To see the effect of varying~$\beta$, first note that
$\beta=0$ corresponds to no measurement of the coin.
In this case the last matrix in Eq.~(\ref{eq:W}) corresponds
to perfect coherence because the $\cos\beta\pi/2$ term becomes unity. As $\beta$
increases from $0$ to~$1$, $\cos\beta\pi/2\rightarrow 0$ and perfect decoherence
emerges: the random walk arises because the end states of the quantum walk are
just probability-weighted sums of each path.

\section{Quantum walks in open systems}
\label{sec:nonunitary}

Nonunitary evolution of the quantum walk of course
can also arise from other processes besides measuring the coin;
in the case of the quantum walk on the cycle in a cavity quantum
electrodynamics realization, cavity damping is naturally associated
with measuring the phase of the intracavity field~\cite{sanders02a},
which corresponds to direct measurements of the vertex states for the
quantum walk on the cycle.
In this case the decoherence mechanism is due to measurements of
vertex occupation, not the coin state, but
the classical random walk emerges all the same.

For more general quantum walks,
weak or strong measurements of the state of the coin
generalize the unitary evolution of Eq.~(\ref{T:unitary})
to the nonunitary case
\begin{equation}
	\mathcal{T}:\rho\mapsto\sum_{i\in\Theta}SC_i\rho C_i^\dagger S^\dagger
\label{T:nonunitary}
\end{equation}
with~$C_i$ the Kraus operators for different instances of coin-state 
randomization.

According to Naimark's theorem~\cite{nielsenchuang00}, the positive
operator-valued measure of coin state that yields the CP map of
Eq.~(\ref{T:nonunitary}) can be realized by coupling the coin state in
$\mathcal{H}_\text{c}$ to an ancilla in an extended Hilbert space
$\mathcal{H}_\text{a}$ and performing
projective (von Neumann) measurements of the ancilla state.
The positive operator-valued measure of the coin state is then obtained by tracing over
ancilla states. The coin can be considered as a qudit of
dimension~$d$, and a $d$-dimensional ancillary qudit suffices. The qudit-qudit
coupling that interpolates from no measurement of the coin state to weak
measurements to sharp measurements with full information is a complicated
generalization~\cite{gottesman98a} of the qubit-qubit coupling
in Eq.~(\ref{eq:ndop}).
A projective measurement of the ancilla gives none, some, or all of the
information about the coin qudit, depending on the qudit-qudit coupling
strength, which, with full measurements, then yields the classical walk on
the graph. Measurements of coin states are sufficient to reduce
a quantum walk on the graph to a classical walk on the same graph
provided the walker starts in a basis state~\cite{lopez03a}.

\section{Optical quantum quincunx}
\label{sec:optqquincunx}

Having incorporated measurement into quantum walks and shown that measurement
introduces complementarity, which interpolates between the random walk
and the unitary quantum walk, we now consider the quantum quincunx:
a physical system which implements a quantum walk.
Our requirements for a quantum quincunx are that the system
(i)~has an identifiable walker,
(ii)~exhibits both the unitary quantum walk and the random walk as
complementary features of the quantum walk, and
(iii)~interpolates between these two complementary extremes
according to a controllable degree of measurement that provides
information about the walker's path. 

The quantum quincunx can be realized in various physical systems~\cite{travaglione01a,sanders02a,dur02a}, but here the optical
quincunx provides a convenient system for understanding quincunxes and
the requirements for a quantum quincunx. Furthermore, the optical quincunx
has been realized experimentally as an interferometer \cite{bouwmeester99a}.
The typical source for interferometry is a coherent laser source, which is
often described as producing a coherent state~\cite{jeong03a}, 
a certain coherent superposition of different numbers of photons,
where the photon number states are given by $|n\rangle$, and the
indeterminacy of the photon number is necessary in order for the phase
variance to be small. Although interferometric
experiments can be fully described by classical fields, the coherent
state provides a bridge to connect the quantum and classical field
descriptions, with the photons playing the role of the `walkers' in the system.

Second quantization seems to present a dilemma with respect to the requirement
of an identifiable walker:
a coherent state of walkers is given by
\begin{equation}
	|\alpha\rangle_\text{w}
		= \text{e}^{-|\alpha|^2/2}\sum_{n=0}^{\infty} \frac{\alpha^n}{\sqrt{n!}}|n\rangle_\text{w}~,
\label{coherentstate}
\end{equation}
where~$|n\rangle_\text{w}$ is a number state of walkers along the graph.
The parameter~$\alpha$ is complex, whose square modulus~$|\alpha|^2$
is the mean energy, the discrete energy distribution is Poissonian, 
and arg$(\alpha)$ is the phase.
Let us deal with two challenges: (a)~the quantum walk with multiple walkers
and (b)~the indefiniteness of the number of walkers.
With respect to challenge~(a), generalizing the quantum walk from one to
$n$ walkers is straightforward: as the $n$-walker system involves
non-interacting walkers, so the Hilbert space for the walkers is
given by $\mathcal{H}_\text{vc}^{\otimes n}$,
and the completely positive map generalizes to $\mathcal{T}^{\otimes n}$.
The $n$-walker system is a simple extension of the one-walker system as a tensor
product of $n$ one-walker systems with one time step given by 
$\mathcal{T}^{\otimes n} \rho^{\otimes n}$.
Each walker carries its own coin, and these $n$ coins are coupled to
$n$ meter qudits, so measurement and complementarity arise via this coupling.
Essentially this $n$-walker system is equivalent to repeating the
one-walker quantum walk $n$ times. 

With respect to challenge~(b), the wave walk appears to emerge through
second quantization of walker number, and the indeterminacy of walker number
in Eq.~(\ref{coherentstate}) enables the phase, which is complementary
to number, to be reasonably sharp in order to provide strong coherence.
However, it has become abundantly clear
recently \cite{Molmer97a,Rudolph01a,Sanders03a} that the coherent state and
number state offer complementary yet convenient alternative representations.
The quantum optics implementation \cite{bouwmeester99a,knight03a,knight03b,jeong03a}
can be described within the photon number superselection
framework~\cite{Sanders03a}, so each run of the optical quincunx
experiment can be interpreted as having a fixed number of photons,
and this number of photons can be post selected by an ideal photon
counting measurement on all the output fields.

Complementarity in the quantum walk would be manifested by 
allowing each photon to be tracked during its evolution.
This requirement is not easily met,
but a practical approach is as follows: the parameter $|\alpha|^2$ in
Eq.~(\ref{coherentstate}) corresponds to the laser flux,
and attenuating the laser so that $|\alpha|^2 \ll 1$ ensures
that multiphoton contributions are negligible.
Then each run overwhelmingly corresponds to no photon or one photon.
In this single-photon regime~\cite{Pfleegor67a,Hariharan96a},
the presence of the photon can be ascertained by a photodetection at
the output: the presence of the photon is postselected
when the photodetector at the output clicks, announcing that this run of
the quantum quincunx had a walker.
The photon's path is ascertained by quantum nondemolition
measurements, either by a nonlinear optical medium for deterministic detection
of the photon without destroying it~\cite{Sanders89a} or by nondeterministic
linear optical detection~\cite{pryde04a}.
Operating in a low photon flux regime
and performing photon number quantum nondemolition measurements of path
would allow a continuous interpolation between the 
the unitary quantum walk and the random walk,
thus extending the optical quincunx to a fully functioning quantum quincunx. 
We are thus able to elucidate what has been achieved by 
the optical quincunx of Bouwmeester et al~\cite{bouwmeester99a}
towards an optical implementation of the quantum walk and what needs
to be added to provide a full implementation.

The optical quincunx emulates the undular properties, or interference, of the
quantum walk, as stated by Knight et al~\cite{knight03b}, but this is
performed without an identifiable, single walker.
In fact the transition to the distribution for a classical random walk
should also be achievable if the interferometer is allowed to decohere.
If the relative phases between interferometer paths are fully decohered,
for example by dispersive media in the paths,
the resultant interferometer output will not correspond
to a superposition of paths but rather to an incoherent,
probability-weighted distribution of outputs.
This incoherent sum of paths is precisely the random walk distribution.

Thus, the experiment of Bouwmeester et al could be modified to
demonstrate features of both the random walk distribution and the
quantum walk interference effect, but complementarity dictates three
criteria to achieve quantumness.
Just demonstrating a decoherence of wave like interference is
insufficient to establish the corpuscular property of the objects;
one must demonstrate their indivisibility.
The walker should be a single photon, and there are several possible
methods to achieve this~\cite{Hariharan96a}.
One is to produce photon pairs, for example via parametric downconversion,
and obtain one photon conditioned on detecting its partner;
another approach is to produce single photons on demand
by a source such as a quantum dot in a strong cavity;
a third approach, which is currently the easiest,
is to postselect the results on detecting a single photon,
thereby eliminating vacuum and
multiphoton contributions and sorting data based on one and only
one photon having been in the system. 

In summary, an optical quincunx can implement the quantum
walk when it operates in the single photon regime.
In this case the device is a \emph{quantum} optical 
quincunx. The quantumness of the quincunx is essential
to manifest the complementarity properties of the quantum walk,
namely the trade-off between information about the
walker's path and the interference.

\section{Conclusions}
\label{sec:conc}

In this paper we have incorporated complementarity into the theory
of quantum walks, thereby addressing the issue of what is ``quantum''
about the quantum walk, as well as extending the concept of complementarity 
well beyond the usual physical systems (e.g.~interferometry) to quantum
walks on general graphs.
Our analysis of complementarity in quantum walks builds on the
approach of coined quantum walks and replaces unitary evolution
by the much more general completely positive map approach,
which is relevant to considerations of experimental
realizations of quantum walks.
Through measurement, the quantum walk may exhibit the unitary quantum walk,
the random walk, and intermediate processes depending
on the strength of the measurement.
We define a quantum quincunx as a physical implementation of
a quantum walk, including measurement of path, that can demonstrate
the essential properties of complementarity in a quantum walk:
interference traded off with which-path information, and the indivisibility
of the quantum walker.  An extension of the optical quincunx experiment
of Bouwmeester et al~\cite{bouwmeester99a} operated in the single
photon regime with postselection for the presence of a photon, and 
in which the interferometer arms contained photon nondemolition
measurement devices, would turn this into a fully quantum quincunx.

\begin{acknowledgments}

VK appreciates useful discussions 
with P.\ L.\ Knight, E.\ Rold{\'a}n, and J.\ Sipe.
BCS acknowledges valuable discussions with S.\ D.\ Bartlett,
D.\ W.\ Berry, M.\ Hillery, D.\ Meyer and J.\ Watrous.
This work was funded in part by the UK Engineering and Physical Sciences
Research Council grant number GR/N2507701,
Alberta's informatics Circle of Research Excellence (iCORE), and
the Australian DEST
Innovation Access Program fund to support collaboration with
the European Fifth Framework project QUPRODIS.

\end{acknowledgments}

\bibliography{qrw,qit,qop}

\begin{thebibliography}{37}
\expandafter\ifx\csname natexlab\endcsname\relax\def\natexlab#1{#1}\fi
\expandafter\ifx\csname bibnamefont\endcsname\relax
  \def\bibnamefont#1{#1}\fi
\expandafter\ifx\csname bibfnamefont\endcsname\relax
  \def\bibfnamefont#1{#1}\fi
\expandafter\ifx\csname citenamefont\endcsname\relax
  \def\citenamefont#1{#1}\fi
\expandafter\ifx\csname url\endcsname\relax
  \def\url#1{\texttt{#1}}\fi
\expandafter\ifx\csname urlprefix\endcsname\relax\def\urlprefix{URL }\fi
\providecommand{\bibinfo}[2]{#2}
\providecommand{\eprint}[2][]{\url{#2}}

\bibitem[{\citenamefont{Farhi and Gutmann}(1998)}]{farhi98a}
\bibinfo{author}{\bibfnamefont{E.}~\bibnamefont{Farhi}} \bibnamefont{and}
  \bibinfo{author}{\bibfnamefont{S.}~\bibnamefont{Gutmann}},
  \bibinfo{journal}{Phys.~Rev.~A} \textbf{\bibinfo{volume}{58}},
  \bibinfo{pages}{915} (\bibinfo{year}{1998}), \eprint{quant-ph/9706062}.

\bibitem[{\citenamefont{Watrous}(2001)}]{watrous98a}
\bibinfo{author}{\bibfnamefont{J.}~\bibnamefont{Watrous}}, \bibinfo{journal}{J.
  Comp. System Sciences} \textbf{\bibinfo{volume}{62}}, \bibinfo{pages}{376}
  (\bibinfo{year}{2001}), \eprint{cs.CC/9812012}.

\bibitem[{\citenamefont{Ambainis et~al.}(2001)\citenamefont{Ambainis, Bach,
  Nayak, Vishwanath, and Watrous}}]{ambainis01a}
\bibinfo{author}{\bibfnamefont{A.}~\bibnamefont{Ambainis}},
  \bibinfo{author}{\bibfnamefont{E.}~\bibnamefont{Bach}},
  \bibinfo{author}{\bibfnamefont{A.}~\bibnamefont{Nayak}},
  \bibinfo{author}{\bibfnamefont{A.}~\bibnamefont{Vishwanath}},
  \bibnamefont{and} \bibinfo{author}{\bibfnamefont{J.}~\bibnamefont{Watrous}},
  in \emph{\bibinfo{booktitle}{Proc.~33rd Annual ACM Symposium on Theory of
  Computing (STOC 2001)}} (\bibinfo{publisher}{Assoc. for Comp. Machinery, New
  York}, \bibinfo{year}{2001}), pp. \bibinfo{pages}{60--69}.

\bibitem[{\citenamefont{Aharanov et~al.}(2001)\citenamefont{Aharanov, Ambainis,
  Kempe, and Vazirani}}]{aharonov00a}
\bibinfo{author}{\bibfnamefont{D.}~\bibnamefont{Aharanov}},
  \bibinfo{author}{\bibfnamefont{A.}~\bibnamefont{Ambainis}},
  \bibinfo{author}{\bibfnamefont{J.}~\bibnamefont{Kempe}}, \bibnamefont{and}
  \bibinfo{author}{\bibfnamefont{U.}~\bibnamefont{Vazirani}}, in
  \emph{\bibinfo{booktitle}{Proc.~33rd Annual ACM Symposium on Theory of
  Computing (STOC 2001)}} (\bibinfo{publisher}{Assoc. for Comp. Machinery, New
  York}, \bibinfo{year}{2001}), pp. \bibinfo{pages}{50--59},
  \eprint{quant-ph/0012090}.

\bibitem[{\citenamefont{Travaglione and Milburn}(2002)}]{travaglione01a}
\bibinfo{author}{\bibfnamefont{B.~C.} \bibnamefont{Travaglione}}
  \bibnamefont{and} \bibinfo{author}{\bibfnamefont{G.~J.}
  \bibnamefont{Milburn}}, \bibinfo{journal}{Phys.~Rev.~A}
  \textbf{\bibinfo{volume}{65}}, \bibinfo{pages}{032310}
  (\bibinfo{year}{2002}), \eprint{quant-ph/0109076}.

\bibitem[{\citenamefont{Sanders
  et~al.}(2003{\natexlab{a}})\citenamefont{Sanders, Bartlett, Tregenna, and
  Knight}}]{sanders02a}
\bibinfo{author}{\bibfnamefont{B.~C.} \bibnamefont{Sanders}},
  \bibinfo{author}{\bibfnamefont{S.~D.} \bibnamefont{Bartlett}},
  \bibinfo{author}{\bibfnamefont{B.}~\bibnamefont{Tregenna}}, \bibnamefont{and}
  \bibinfo{author}{\bibfnamefont{P.~L.} \bibnamefont{Knight}},
  \bibinfo{journal}{Phys.~Rev.~A} \textbf{\bibinfo{volume}{67}},
  \bibinfo{pages}{042305} (\bibinfo{year}{2003}{\natexlab{a}}),
  \eprint{quant-ph/0207028}.

\bibitem[{\citenamefont{{D\"{u}r} et~al.}(2002)\citenamefont{{D\"{u}r},
  Raussendorf, Kendon, and Briegel}}]{dur02a}
\bibinfo{author}{\bibfnamefont{W.}~\bibnamefont{{D\"{u}r}}},
  \bibinfo{author}{\bibfnamefont{R.}~\bibnamefont{Raussendorf}},
  \bibinfo{author}{\bibfnamefont{V.~M.} \bibnamefont{Kendon}},
  \bibnamefont{and} \bibinfo{author}{\bibfnamefont{H.-J.}
  \bibnamefont{Briegel}}, \bibinfo{journal}{Phys.~Rev.~A}
  \textbf{\bibinfo{volume}{66}}, \bibinfo{pages}{052319}
  (\bibinfo{year}{2002}), \eprint{quant-ph/0207137}.

\bibitem[{\citenamefont{Childs et~al.}(2003)\citenamefont{Childs, Cleve,
  Deotto, Farhi, Gutmann, and Spielman}}]{childs02a}
\bibinfo{author}{\bibfnamefont{A.~M.} \bibnamefont{Childs}},
  \bibinfo{author}{\bibfnamefont{R.}~\bibnamefont{Cleve}},
  \bibinfo{author}{\bibfnamefont{E.}~\bibnamefont{Deotto}},
  \bibinfo{author}{\bibfnamefont{E.}~\bibnamefont{Farhi}},
  \bibinfo{author}{\bibfnamefont{S.}~\bibnamefont{Gutmann}}, \bibnamefont{and}
  \bibinfo{author}{\bibfnamefont{D.~A.} \bibnamefont{Spielman}}, in
  \emph{\bibinfo{booktitle}{Proc.~35th Annual ACM Symposium on Theory of
  Computing (STOC 2003)}} (\bibinfo{publisher}{Assoc.~for Comp.~Machinery, New
  York}, \bibinfo{year}{2003}), pp. \bibinfo{pages}{59--68},
  \eprint{quant-ph/0209131}.

\bibitem[{\citenamefont{Shenvi et~al.}(2003)\citenamefont{Shenvi, Kempe, and
  {Birgitta Whaley}}}]{shenvi02a}
\bibinfo{author}{\bibfnamefont{N.}~\bibnamefont{Shenvi}},
  \bibinfo{author}{\bibfnamefont{J.}~\bibnamefont{Kempe}}, \bibnamefont{and}
  \bibinfo{author}{\bibfnamefont{K.}~\bibnamefont{{Birgitta Whaley}}},
  \bibinfo{journal}{Phys.~Rev.~A} \textbf{\bibinfo{volume}{67}},
  \bibinfo{pages}{052307} (\bibinfo{year}{2003}), \eprint{quant-ph/0210064}.

\bibitem[{\citenamefont{Ambainis}(2003)}]{ambainis04a}
\bibinfo{author}{\bibfnamefont{A.}~\bibnamefont{Ambainis}},
  \bibinfo{journal}{Intl.~J.~ Quantum Information}
  \textbf{\bibinfo{volume}{1}}, \bibinfo{pages}{507} (\bibinfo{year}{2003}),
  \eprint{quant-ph/0403120}.

\bibitem[{\citenamefont{Bouwmeester et~al.}(1999)\citenamefont{Bouwmeester,
  Marzoli, Karman, Schleich, and Woerdman}}]{bouwmeester99a}
\bibinfo{author}{\bibfnamefont{D.}~\bibnamefont{Bouwmeester}},
  \bibinfo{author}{\bibfnamefont{I.}~\bibnamefont{Marzoli}},
  \bibinfo{author}{\bibfnamefont{G.~P.} \bibnamefont{Karman}},
  \bibinfo{author}{\bibfnamefont{W.}~\bibnamefont{Schleich}}, \bibnamefont{and}
  \bibinfo{author}{\bibfnamefont{J.~P.} \bibnamefont{Woerdman}},
  \bibinfo{journal}{Phys.~Rev.~A} \textbf{\bibinfo{volume}{61}},
  \bibinfo{pages}{013410} (\bibinfo{year}{1999}).

\bibitem[{\citenamefont{Knight et~al.}(2003{\natexlab{a}})\citenamefont{Knight,
  Rold{\'a}n, and Sipe}}]{knight03a}
\bibinfo{author}{\bibfnamefont{P.~L.} \bibnamefont{Knight}},
  \bibinfo{author}{\bibfnamefont{E.}~\bibnamefont{Rold{\'a}n}},
  \bibnamefont{and} \bibinfo{author}{\bibfnamefont{J.~E.} \bibnamefont{Sipe}},
  \bibinfo{journal}{Phys.~Rev.~A} \textbf{\bibinfo{volume}{68}},
  \bibinfo{pages}{020301} (\bibinfo{year}{2003}{\natexlab{a}}),
  \eprint{quant-ph/0304201}.

\bibitem[{\citenamefont{Knight et~al.}(2003{\natexlab{b}})\citenamefont{Knight,
  Rold{\'a}n, and Sipe}}]{knight03b}
\bibinfo{author}{\bibfnamefont{P.~L.} \bibnamefont{Knight}},
  \bibinfo{author}{\bibfnamefont{E.}~\bibnamefont{Rold{\'a}n}},
  \bibnamefont{and} \bibinfo{author}{\bibfnamefont{J.~E.} \bibnamefont{Sipe}},
  \bibinfo{journal}{Optics Comms.} \textbf{\bibinfo{volume}{227}},
  \bibinfo{pages}{147} (\bibinfo{year}{2003}{\natexlab{b}}),
  \eprint{quant-ph/0305165}.

\bibitem[{\citenamefont{Jeong et~al.}(2004)\citenamefont{Jeong, Paternostro,
  and Kim}}]{jeong03a}
\bibinfo{author}{\bibfnamefont{H.}~\bibnamefont{Jeong}},
  \bibinfo{author}{\bibfnamefont{M.}~\bibnamefont{Paternostro}},
  \bibnamefont{and} \bibinfo{author}{\bibfnamefont{M.~S.} \bibnamefont{Kim}},
  \bibinfo{journal}{Phys.~Rev.~A} \textbf{\bibinfo{volume}{69}},
  \bibinfo{pages}{012310} (\bibinfo{year}{2004}), \eprint{quant-ph/0305008}.

\bibitem[{\citenamefont{Bohr}(1950)}]{bohr50a}
\bibinfo{author}{\bibfnamefont{N.}~\bibnamefont{Bohr}},
  \bibinfo{journal}{Science} \textbf{\bibinfo{volume}{111}},
  \bibinfo{pages}{51} (\bibinfo{year}{1950}), \bibinfo{note}{reprinted from
  Dialectica 2 (1948), 312--319}.

\bibitem[{\citenamefont{Bohr}(1928{\natexlab{a}})}]{bohr28a}
\bibinfo{author}{\bibfnamefont{N.}~\bibnamefont{Bohr}},
  \bibinfo{journal}{Naturwissenschaften} \textbf{\bibinfo{volume}{16}},
  \bibinfo{pages}{245} (\bibinfo{year}{1928}{\natexlab{a}}).

\bibitem[{\citenamefont{Bohr}(1928{\natexlab{b}})}]{bohr28b}
\bibinfo{author}{\bibfnamefont{N.}~\bibnamefont{Bohr}},
  \bibinfo{journal}{Nature (London)} \textbf{\bibinfo{volume}{121}},
  \bibinfo{pages}{580} (\bibinfo{year}{1928}{\natexlab{b}}).

\bibitem[{\citenamefont{Wootters and Zurek}(1979)}]{wootters79a}
\bibinfo{author}{\bibfnamefont{W.~K.} \bibnamefont{Wootters}} \bibnamefont{and}
  \bibinfo{author}{\bibfnamefont{W.~H.} \bibnamefont{Zurek}},
  \bibinfo{journal}{Phys.~Rev.~D} \textbf{\bibinfo{volume}{19}},
  \bibinfo{pages}{473} (\bibinfo{year}{1979}).

\bibitem[{\citenamefont{Sanders and Milburn}(1989)}]{Sanders89a}
\bibinfo{author}{\bibfnamefont{B.~C.} \bibnamefont{Sanders}} \bibnamefont{and}
  \bibinfo{author}{\bibfnamefont{G.~J.} \bibnamefont{Milburn}},
  \bibinfo{journal}{Phys.~Rev.~A} \textbf{\bibinfo{volume}{39}},
  \bibinfo{pages}{694} (\bibinfo{year}{1989}).

\bibitem[{\citenamefont{Englert}(1996)}]{englert96}
\bibinfo{author}{\bibfnamefont{B.-G.} \bibnamefont{Englert}},
  \bibinfo{journal}{Phys.~Rev.~Lett.} \textbf{\bibinfo{volume}{77}},
  \bibinfo{pages}{2154} (\bibinfo{year}{1996}).

\bibitem[{\citenamefont{Pryde et~al.}(2004)\citenamefont{Pryde, O'Brien, White,
  Bartlett, and Ralph}}]{pryde04a}
\bibinfo{author}{\bibfnamefont{G.~J.} \bibnamefont{Pryde}},
  \bibinfo{author}{\bibfnamefont{J.~L.} \bibnamefont{O'Brien}},
  \bibinfo{author}{\bibfnamefont{A.~G.} \bibnamefont{White}},
  \bibinfo{author}{\bibfnamefont{S.~D.} \bibnamefont{Bartlett}},
  \bibnamefont{and} \bibinfo{author}{\bibfnamefont{T.~C.} \bibnamefont{Ralph}},
  \bibinfo{journal}{Phys.~Rev.~Lett.} \textbf{\bibinfo{volume}{92}},
  \bibinfo{pages}{190402} (\bibinfo{year}{2004}).

\bibitem[{\citenamefont{Moore and Russell}(2002)}]{moore01a}
\bibinfo{author}{\bibfnamefont{C.}~\bibnamefont{Moore}} \bibnamefont{and}
  \bibinfo{author}{\bibfnamefont{A.}~\bibnamefont{Russell}}, in
  \emph{\bibinfo{booktitle}{Proc.~6th Intl.~Workshop on Randomization and
  Approximation Techniques in Computer Science (RANDOM '02)}}, edited by
  \bibinfo{editor}{\bibfnamefont{J.~D.~P.} \bibnamefont{Rolim}}
  \bibnamefont{and} \bibinfo{editor}{\bibfnamefont{S.}~\bibnamefont{Vadhan}}
  (\bibinfo{publisher}{Springer}, \bibinfo{year}{2002}), pp.
  \bibinfo{pages}{164--178}, \eprint{quant-ph/0104137}.

\bibitem[{\citenamefont{Mackay et~al.}(2002)\citenamefont{Mackay, Bartlett,
  Stephenson, and Sanders}}]{mackay01a}
\bibinfo{author}{\bibfnamefont{T.~D.} \bibnamefont{Mackay}},
  \bibinfo{author}{\bibfnamefont{S.~D.} \bibnamefont{Bartlett}},
  \bibinfo{author}{\bibfnamefont{L.~T.} \bibnamefont{Stephenson}},
  \bibnamefont{and} \bibinfo{author}{\bibfnamefont{B.~C.}
  \bibnamefont{Sanders}}, \bibinfo{journal}{J.~Phys.~A: Math.~Gen.}
  \textbf{\bibinfo{volume}{35}}, \bibinfo{pages}{2745} (\bibinfo{year}{2002}),
  \eprint{quant-ph/0108004}.

\bibitem[{\citenamefont{Hillery et~al.}(2003)\citenamefont{Hillery, Bergou, and
  Feldman}}]{hillery03a}
\bibinfo{author}{\bibfnamefont{M.}~\bibnamefont{Hillery}},
  \bibinfo{author}{\bibfnamefont{J.}~\bibnamefont{Bergou}}, \bibnamefont{and}
  \bibinfo{author}{\bibfnamefont{E.}~\bibnamefont{Feldman}},
  \bibinfo{journal}{Phys.~Rev.\ A} \textbf{\bibinfo{volume}{68}},
  \bibinfo{pages}{032314} (\bibinfo{year}{2003}), \eprint{quant-ph/0302161}.

\bibitem[{\citenamefont{Aharonov et~al.}(1992)\citenamefont{Aharonov,
  Davidovich, and Zagury}}]{aharonov92a}
\bibinfo{author}{\bibfnamefont{Y.}~\bibnamefont{Aharonov}},
  \bibinfo{author}{\bibfnamefont{L.}~\bibnamefont{Davidovich}},
  \bibnamefont{and} \bibinfo{author}{\bibfnamefont{N.}~\bibnamefont{Zagury}},
  \bibinfo{journal}{Phys.~Rev.~A} \textbf{\bibinfo{volume}{48}},
  \bibinfo{pages}{1687} (\bibinfo{year}{1992}).

\bibitem[{\citenamefont{Kendon and Tregenna}(2003)}]{kendon02c}
\bibinfo{author}{\bibfnamefont{V.}~\bibnamefont{Kendon}} \bibnamefont{and}
  \bibinfo{author}{\bibfnamefont{B.}~\bibnamefont{Tregenna}},
  \bibinfo{journal}{Phys.~Rev.~A} \textbf{\bibinfo{volume}{67}},
  \bibinfo{pages}{042315} (\bibinfo{year}{2003}), \eprint{quant-ph/0209005}.

\bibitem[{\citenamefont{Berry}(2003)}]{berry03a}
\bibinfo{author}{\bibfnamefont{D.~W.} \bibnamefont{Berry}}
  (\bibinfo{year}{2003}), \bibinfo{note}{private communication}.

\bibitem[{\citenamefont{Zurek}(1981)}]{zurek81a}
\bibinfo{author}{\bibfnamefont{W.~H.} \bibnamefont{Zurek}},
  \bibinfo{journal}{Phys.~Rev.~D} \textbf{\bibinfo{volume}{24}},
  \bibinfo{pages}{1516} (\bibinfo{year}{1981}).

\bibitem[{\citenamefont{Zurek}(1982)}]{zurek82a}
\bibinfo{author}{\bibfnamefont{W.~H.} \bibnamefont{Zurek}},
  \bibinfo{journal}{Phys.~Rev.~D} \textbf{\bibinfo{volume}{26}},
  \bibinfo{pages}{1862} (\bibinfo{year}{1982}).

\bibitem[{\citenamefont{Nielsen and Chuang}(2000)}]{nielsenchuang00}
\bibinfo{author}{\bibfnamefont{M.~A.} \bibnamefont{Nielsen}} \bibnamefont{and}
  \bibinfo{author}{\bibfnamefont{I.~J.} \bibnamefont{Chuang}},
  \emph{\bibinfo{title}{Quantum Computation and Quantum Information}}
  (\bibinfo{publisher}{Cambridge University Press, Cambs. UK},
  \bibinfo{year}{2000}).

\bibitem[{\citenamefont{Gottesman}(1999)}]{gottesman98a}
\bibinfo{author}{\bibfnamefont{D.}~\bibnamefont{Gottesman}},
  \bibinfo{journal}{Lecture Notes in Comp.~Sci.}
  \textbf{\bibinfo{volume}{1509}}, \bibinfo{pages}{302} (\bibinfo{year}{1999}),
  \eprint{quant-ph/9802007}.

\bibitem[{\citenamefont{Lop\'{e}z and Paz}(2003)}]{lopez03a}
\bibinfo{author}{\bibfnamefont{C.~C.} \bibnamefont{Lop\'{e}z}}
  \bibnamefont{and} \bibinfo{author}{\bibfnamefont{J.~P.} \bibnamefont{Paz}},
  \bibinfo{journal}{Phys.~Rev.~A} \textbf{\bibinfo{volume}{68}},
  \bibinfo{pages}{052305} (\bibinfo{year}{2003}), \eprint{quant-ph/0308104}.

\bibitem[{\citenamefont{{M{\o}lmer}}(1997)}]{Molmer97a}
\bibinfo{author}{\bibfnamefont{K.}~\bibnamefont{{M{\o}lmer}}},
  \bibinfo{journal}{Phys.~Rev.~A} \textbf{\bibinfo{volume}{55}},
  \bibinfo{pages}{3195} (\bibinfo{year}{1997}).

\bibitem[{\citenamefont{Rudolph and Sanders}(2001)}]{Rudolph01a}
\bibinfo{author}{\bibfnamefont{T.}~\bibnamefont{Rudolph}} \bibnamefont{and}
  \bibinfo{author}{\bibfnamefont{B.~C.} \bibnamefont{Sanders}},
  \bibinfo{journal}{Phys.~Rev.~Lett.} \textbf{\bibinfo{volume}{87}},
  \bibinfo{pages}{077903} (\bibinfo{year}{2001}).

\bibitem[{\citenamefont{Sanders
  et~al.}(2003{\natexlab{b}})\citenamefont{Sanders, Bartlett, Rudolph, and
  Knight}}]{Sanders03a}
\bibinfo{author}{\bibfnamefont{B.~C.} \bibnamefont{Sanders}},
  \bibinfo{author}{\bibfnamefont{S.~D.} \bibnamefont{Bartlett}},
  \bibinfo{author}{\bibfnamefont{T.}~\bibnamefont{Rudolph}}, \bibnamefont{and}
  \bibinfo{author}{\bibfnamefont{P.~L.} \bibnamefont{Knight}},
  \bibinfo{journal}{Phys.~Rev.~A} \textbf{\bibinfo{volume}{68}},
  \bibinfo{pages}{042329} (\bibinfo{year}{2003}{\natexlab{b}}).

\bibitem[{\citenamefont{Pfleegor and Mandel}(1967)}]{Pfleegor67a}
\bibinfo{author}{\bibfnamefont{P.}~\bibnamefont{Pfleegor}} \bibnamefont{and}
  \bibinfo{author}{\bibfnamefont{L.}~\bibnamefont{Mandel}},
  \bibinfo{journal}{Phys.~Rev.} \textbf{\bibinfo{volume}{159}},
  \bibinfo{pages}{1084} (\bibinfo{year}{1967}).

\bibitem[{\citenamefont{Hariharan and Sanders}(1996)}]{Hariharan96a}
\bibinfo{author}{\bibfnamefont{P.}~\bibnamefont{Hariharan}} \bibnamefont{and}
  \bibinfo{author}{\bibfnamefont{B.~C.} \bibnamefont{Sanders}},
  \bibinfo{journal}{Prog.~Opt.} \textbf{\bibinfo{volume}{36}},
  \bibinfo{pages}{49} (\bibinfo{year}{1996}).

\end{thebibliography}

\end{document}